\documentclass[a4paper,showpacs,prb,twocolumn]{revtex4}
\usepackage{amsfonts}
\usepackage{graphicx}
\usepackage{color}
\usepackage{amsmath}
\usepackage{amssymb}
\usepackage{latexsym}
\usepackage{psfrag}

\begin{document}

\title{Magnetoconductance of interacting electrons in quantum wires: Spin
density functional theory study}
\author{S. Ihnatsenka}
\affiliation{Solid State Electronics, Department of Science and Technology (ITN), Link%
\"{o}ping University, 60174 Norrk\"{o}ping, Sweden}
\date{\today }
\author{I. V. Zozoulenko}
\affiliation{Solid State Electronics, Department of Science and Technology (ITN), Link%
\"{o}ping University, 60174 Norrk\"{o}ping, Sweden}
\date{\today }

\begin{abstract}
We present systematic quantitative description of the magnetoconductance of
split-gate quantum wires focusing on formation and evolution of the odd
(spin-resolved) conductance plateaus. We start from the case of spinless
electrons where the calculated magnetoconductance in the Hartree
approximation shows the plateaus quantized in units of 2$e^{2}/h$ separated
by transition regions whose width grows as the magnetic field is increased.
We show that the transition regions are related to the formation of the
compressible strips in the middle of the wire occupied by electrons
belonging to the highest (spin-degenerate) subband. Accounting for the
exchange and correlation interactions within the spin density functional
theory (DFT) leads to lifting of the spin degeneracy and formation of the
spin-resolved plateaus at odd values of $e^{2}/h.$ The most striking feature
of the magnetoconductance is that the width of the odd conductance steps in
the spin DFT calculations is equal to the width of the transition intervals
between the conductance steps in the Hartree calculations. A detailed
analysis of the evolution of the Hartree and the spin-DFT subband structure
provides an explanation of this finding. Our calculations also reveal the
effect of the collapse of the odd conductance plateaus for lower fields. We
attribute this effect to the reduced screening efficiency in the confined
(wire) geometry when the width of the compressible strip in the center
becomes much smaller than the extend of the wave function. A detailed
comparison to the experimental data demonstrates that the spin-DFT
calculations reproduce not only qualitatively, but rather quantitatively all
the features observed in the experiment. This includes the dependence of the
width of the odd and even plateaus on the magnetic field as well as the
estimation of the subband index corresponding to the last resolved odd
plateau in the magnetoconductance.
\end{abstract}

\maketitle

\section{Introduction}

The quantized conductance of a two-dimensional electron gas (2DEG) in the
quantum Hall regime has generated a tremendous attention since its discovery
in 1980\cite{vonKlitzing}. For a theoretical description of the integer
quantum Hall (IQH) effect, the concept of edge states combined with the
Landauer-Buttiker formalism is widely used\cite{edges,BvH}. This approach is
proved to be especially appealing for the description of electron transport
in the quantum Hall regime in confined geometries like quantum wires or
quantum point contacts (QPCs)\cite{vanWees,BvH}.

Some aspects of the quantized conductance in the confined geometries in the
IQH regime can be understood in a one-electron picture. This includes, for
example, magnetic depopulation of the subbands in a quantum wire\cite%
{Berggren}, and selective population and detection of edge channels by QPCs
resulting in the observation of anomalous IQH effect\cite{vanWees,BvH}. In
the one-electron picture the two-terminal magnetoconductance $G$ of a
quantum wire or a QPC exhibits quantized plateaus in units of $2e^{2}/h$
(for spinless electrons) separated by transition regions of an essentially
zero width. The experiments, however, show that an extend of these
transition regions can be comparable to the width of the plateaus\cite%
{BvH,vanWees,Wrobel,experiment}. This indicates that an accurate description
of the magnetoconductance in the IQH regime even without accounting for spin
effects requires approaches that go beyond a simple one-electron picture of
non-interacting electrons. A quantitative electrostatic theory of
interacting electrons in quantum wires was proposed by Chklovskii, Matveev
and Shklovskii\cite{ChklovskiiII}. They demonstrated that in a strong
magnetic field, alternating strips of compressible and incompressible
liquids are formed in the center of the wire. They also evaluated the
two-terminal magnetoconductance of the wire. In contrast to the one-electron
description, the magnetoconductance of interacting electron was shown to
exhibits very narrow quantized plateaus separated by much broader rises
where the conductance was not quantized. This conclusion (being opposite to
the prediction of the one-electron picture) is also in apparent disagreement
with the experiments. This indicates that even for spinless electrons in the
IQH regime an accurate quantitative description of the magnetoconductance
requires many-body quantum mechanical treatment.

At low temperature in clean high-mobility samples the spin degeneracy is
lifted and the additional plateaus at odd values of $e^{2}/h$ become resolved%
\cite{BvH,experiment}. This is due to the exchange and correlations effects
leading to the strong enhancement of the electron $g$ factor above its bulk
value\cite{Uemura}. The effect of the many-body interactions on the spin
splitting in quantum wires in the IQH regime has been a subject of numerous
studies\cite%
{Kinaret,Dempsey,Tokura,Manolescu,Takis,Takis2,Stoof,Kramer,Studart,Ihnatsenka,Ihnatsenka2,Ihnatsenka3,hysteresis,HF}%
. These studies have focused on various aspects of 2DEG in confined
geometries including the structure of compressible/incompressible strips,
suppression or enhancement of the $g$ factor, subband spin splitting,
spatial spin separation and other. We however have not been able to find in
the literature any systematic quantitative treatment of the
magnetoconductance of the structures at hand. Surprisingly enough, even
after two decades of the studies of IQH systems in the confined geometry,
the question of the fundamental importance addressing the formation of the
odd plateaus in the magnetoconductance and corresponding quantitative
description of the plateau widths still remains unanswered. As discussed
above, the structure of the magnetoconductance plateaus remains poorly
understood even for the case of spinless electrons. Recent advances in the
field such as demonstration of the Mach--Zehnder\cite{Ji}, Aharonov-Bohm\cite%
{Goldman_AB} , and Laughlin quasiparticle interferometers\cite{Goldman} or
prospects of the topological quantum computing\cite{TopQuantComp} has led to
a renewed interested to the magnetoconductance in the quantum wires and
related structures. Even though many of the the above systems operate in the
fractional quantum Hall regime where the correlation effects become
dominant, a detailed understanding of the magnetoconductance in the IQH
regime is the necessary prerequisite for the understanding of the
magnetotransport in the fractional regime.

In our previous publications we provided a systematic quantitative
description of the structure and spin polarization of edge states and
magnetosubband evolution in the quantum wire based on the self-consistent
Green's function techniques combined with the spin density functional theory
(DFT)\cite{Ihnatsenka,Ihnatsenka2,Ihnatsenka3} or Hartree-Fock approach\cite%
{HF}. The main aim of this paper is to present a systematic quantitative
description of the two-terminal magnetonductance of the quantum wire with
the focus on the formation and evolution of the exchange-induced odd
conductance plateaus. The motivation for the present paper is the recent
experimental studies of the spin-resolved magnetoconductance of the narrow
channels in the IQH regime\cite{experiment}. One of the remarkable finding
of this experiment is the collapse of the spin splitting in the confined
geometries for lower field. The spin-DFT magnetotransport calculations
presented in this paper not only capture essential features observed in the
experiment, but demonstrate rather good quantitative agreement with the
calculated and observed magnetoconductances. This includes the width and the
position of the magnetoconductance plateaus (both odd and even), as well as
predictions for the critical magnetic field where the odd plateaus disappear
in the magnetoconductance. We therefore conclude that the spin DFT approach
represents the powerful tool to study large realistic quantum Hall systems
containing hundreds or thousands of electrons, providing detailed and
reliable microscopic information on wavefunctions, electron densities and
currents as well as the conductance.

\section{Basics}

\begin{figure}[tb]
\includegraphics[scale=1.0]{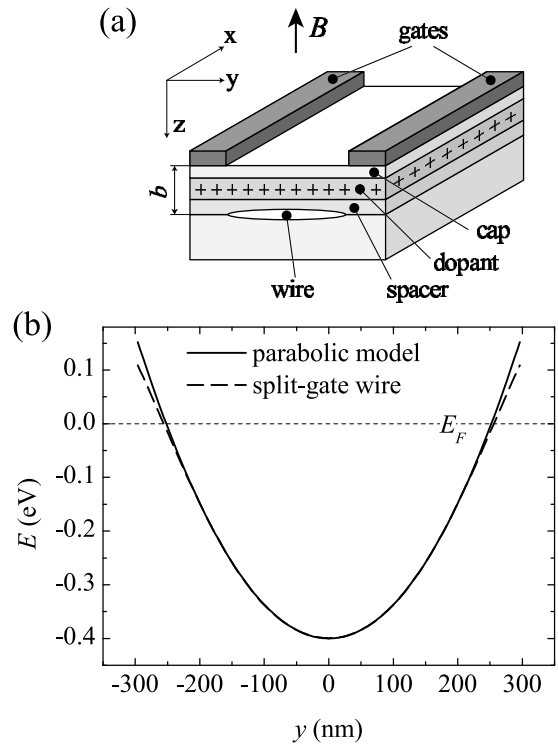} 
\caption{(a) Schematic diagram of a split-gate quantum wire. (b) The bare
electrostatic parabolic confining potential $V_{conf}(y)$ (Eq. (\protect\ref%
{V_conf})) with $V_{0}=-0.4$ eV, $\hbar \protect\omega _{0}=3.8$ meV (solid
lines). Dashed lines shows the calculated electrostatic confinement for a
realistic split-gate quantum wires depicted in (a) with the distance between
the gates $a=500$ nm, the gate voltage $V_{g}=-2$ V, $\protect\delta -$donor
concentration $n_{d}=6\cdot 10^{23}$ m$^{-3}$, the donor and electron
distance to the surface are 100 nm and 200 nm respectively and the Schottky
barrier $V_{Schottky}=0.8$ eV (The potentials from the gates and the donors
are calculated on the basis of Eqs. (2) and (3) of Ref.
\onlinecite{Ihnatsenka}). }
\label{fig:structure}
\end{figure}
We consider an infinitely long wire in a perpendicular magnetic field $B$,
see Fig. \ref{fig:structure}. The bare electrostatic confinement (due to the
split gates, the donor layer and the Schottky barrier) can be approximated
by a parabolic potential,
\begin{equation}
V_{conf}(y)=V_{0}+\frac{m^{\ast }}{2}\left( \omega _{0}y\right) ^{2},
\label{V_conf}
\end{equation}%
where $V_{0}$ is the bottom of the potential, $\omega _{0}$ defines the
potential slope, and $m^{\ast }=0.067m_{e}$ is the electron effective mass
in GaAs. [The comparison of the model parabolic potential with the
calculated potential in a realistic split gate wire is shown in Fig. 1(b)].
By varying $V_{0}$ and $\omega _{0}$ we can change the wire width and the
electron density; in our calculations we set the Fermi energy $E_{F}=0$.

In order to calculate the magnetoconductance of the quantum wire, its
subband structure and the wavefunctions we use the Green's function technique%
\cite{Ihnatsenka,Ihnatsenka2} where the electron interaction and the spin
effects are included self-consistently within the framework of the Kohn-Sham
density function theory in the local spin density approximation\cite%
{Giuliani_Vignale}. (The reliability of the the spin density functional
theory for electronic structure and magnetotransport calculations in quantum
wires, dots and related structures is discussed in details in Refs. \onlinecite%
{HF,conductance}).

We start from the Hamiltonian $H=\sum_{\sigma }\left( H_{0}+V^{\sigma
}(y)\right) ,$ where $H_{0}$ is the kinetic energy in the Landau gauge,
\begin{equation}
H_{0}=-\frac{\hbar ^{2}}{2m^{\ast }}\left\{ \left( \frac{\partial }{\partial
x}-\frac{eiBy}{\hbar }\right) ^{2}+\frac{\partial ^{2}}{\partial y^{2}}%
\right\} ,  \label{H_0}
\end{equation}%
and the total confining potential $V^{\sigma }(y),$
\begin{equation}
V^{\sigma }(y)=V_{conf}(y)+V_{H}(y)+V_{xc}^{\sigma }(y)+V_{Z}^{\sigma }
\label{V_tot}
\end{equation}%
includes the bare electrostatic confinement $V_{conf}(y)$ (given by Eq. (\ref%
{V_conf})), the Hartree potential $V_{H}(y),$ the exchange-correlation
potential $V_{xc}^{\sigma }(y),$ and the Zeeman term, $V_{Z}^{\sigma }=g\mu
_{b}B\sigma ,$ where $\sigma =\pm \frac{1}{2}$ describes the spin-up and
spin-down states, $\uparrow $ ,$\downarrow ;$ $\mu _{b}=e\hbar /2m_{e}$ is
the Bohr magneton, and the bulk $g$ factor of GaAs is $g=-0.44.$ The Hartree
potential $V_{H}(y)$ due to the electron density $n(y)=\sum_{\sigma
}n^{\sigma }(y)$ (including the mirror charges) reads\cite{Ihnatsenka}%
\begin{equation}
V_{H}(y)=-\frac{e^{2}}{4\pi \varepsilon _{0}\varepsilon _{r}}\int_{-\infty
}^{+\infty }dy^{\prime }n(y^{\prime })\ln \frac{\left( y-y^{\prime }\right)
^{2}}{\left( y-y^{\prime }\right) ^{2}+4b^{2}},  \label{V_H}
\end{equation}%
with $b$ being the distance from the electron gas to the surface (we choose $%
b=200$ nm). The exchange and correlation potential $V_{xc}(y)$ in the local
spin density approximation is given by the functional derivative%
\begin{equation}
V_{xc}^{\sigma }(y)=\frac{\delta }{\delta n^{\sigma }}\left\{ n\epsilon
_{xc}\left( n,\zeta (y)\right) \right\}
\end{equation}%
where $\zeta (y)=\frac{n^{\uparrow }-n^{\downarrow }}{n^{\uparrow
}+n^{\downarrow }}$ is the local spin-polarization. In the present paper we
use the parameterization of the exchange and correlation energy $\epsilon
_{xc}$ given by Tanatar and Ceperly (TC)\cite{TC}. Note that we also
performed calculations on the basis of the parametrization recently provided
by Attaccalite \textit{et al.} \cite{AMGB} and found only marginal
difference with the results based on the TC functional.

The spin-resolved electron density in the wire can be expressed via the
Green's function $G^{\sigma }(y,y,E)$
\begin{equation}
n^{\sigma }(y)=-\frac{1}{\pi }\text{Im}\left[ \int_{-\infty }^{\infty
}dE\,G^{\sigma }(y,y,E)f_{FD}(E-E_{F})\right] ,  \label{density}
\end{equation}%
where $f_{FD}(E-E_{F})$ is the Fermi-Dirac distribution function. The
Green's function, the Bloch states, the electron and current densities are
calculated self-consistently using the technique described in detail in Ref.
[\onlinecite{Ihnatsenka}]. Knowledge of the wave vectors $k_{\alpha
}^{\sigma }$ for different Bloch states $\alpha $ allows us to recover the
subband structure, i.e. to calculate an overage position $y_{\alpha
}^{\sigma }$ of the wave functions for different modes $\alpha $ for the
given energy $E$ \cite{Davies_book},
\begin{equation}
y_{\alpha }^{\sigma }=\frac{\hbar k_{\alpha }^{\sigma }}{eB}.  \label{x_b}
\end{equation}%
We calculate the spin-resolved conductance of the wire on the basis of the
linear-response Landauer formula,
\begin{equation}
G^{\sigma }=\frac{e^{2}}{h}\sum_{\alpha }\int_{E_{th\,\alpha }^{\sigma
}}^{\infty }dE\left( -\frac{\partial f\left( E-E_{F}\right) }{\partial E}%
\right) ,  \label{G}
\end{equation}%
where summation is performed over all propagating modes $\alpha $ for the
spin $\sigma ,$ with $E_{th\,\alpha }^{\sigma }$ being the propagation
threshold for $\alpha $-th mode. The current density for a mode $\alpha $ is
calculated as \cite{Ihnatsenka}
\begin{equation}
J_{\alpha }^{\sigma }(y)=\frac{e^{2}}{h}V\int dE\frac{j_{\alpha }^{\sigma
}(y,E)}{v_{\alpha }^{\sigma }}\left( -\frac{\partial f\left( E-E_{F}\right)
}{\partial E}\right) ,  \label{J}
\end{equation}%
with $v_{\alpha }^{\sigma }$ and $j_{\alpha }^{\sigma }(y,E)$ being
respectively the group velocity and the quantum-mechanical particle current
density for the state $\alpha $ at the energy $E$, and $V$ being the applied
voltage. All the calculations presented in this paper are performed for the
temperature $T=100$ mK. In order to speed up the calculation we use the
modified Broyden method\cite{Broyden} that allows one to reduce the number
of iterations need to achieve a self-consistent solution from $\sim 2000$ to
only $\sim 50$ .

\section{Result and discussions}

\subsection{Hartree and spin DFT approximations}

We start our analysis of the magnetoconductance and the magnetosubband
structure in quantum wires from the case of the Hartree approximation when
the exchange and correlation interactions are not included in the effective
potential (i.e. when $V_{xc}^{\sigma }(y)$ is set to zero in Eq. (\ref{V_tot}%
)). Note that the total potential $V^{\sigma }(y)$ also includes the Zeeman
term leading to the spin splitting even in the Hartree case. The effect of
the Zeeman term is however negligibly small in the considered field
intervals. We will thus refer to the Hartree case as for the case of
spinless electrons. The results obtained in the Hartree approximation will
provide a basis for understanding of the effect of the exchange and
correlation within the spin DFT approximation.

\begin{figure*}[tbp]
\includegraphics[scale=0.8]{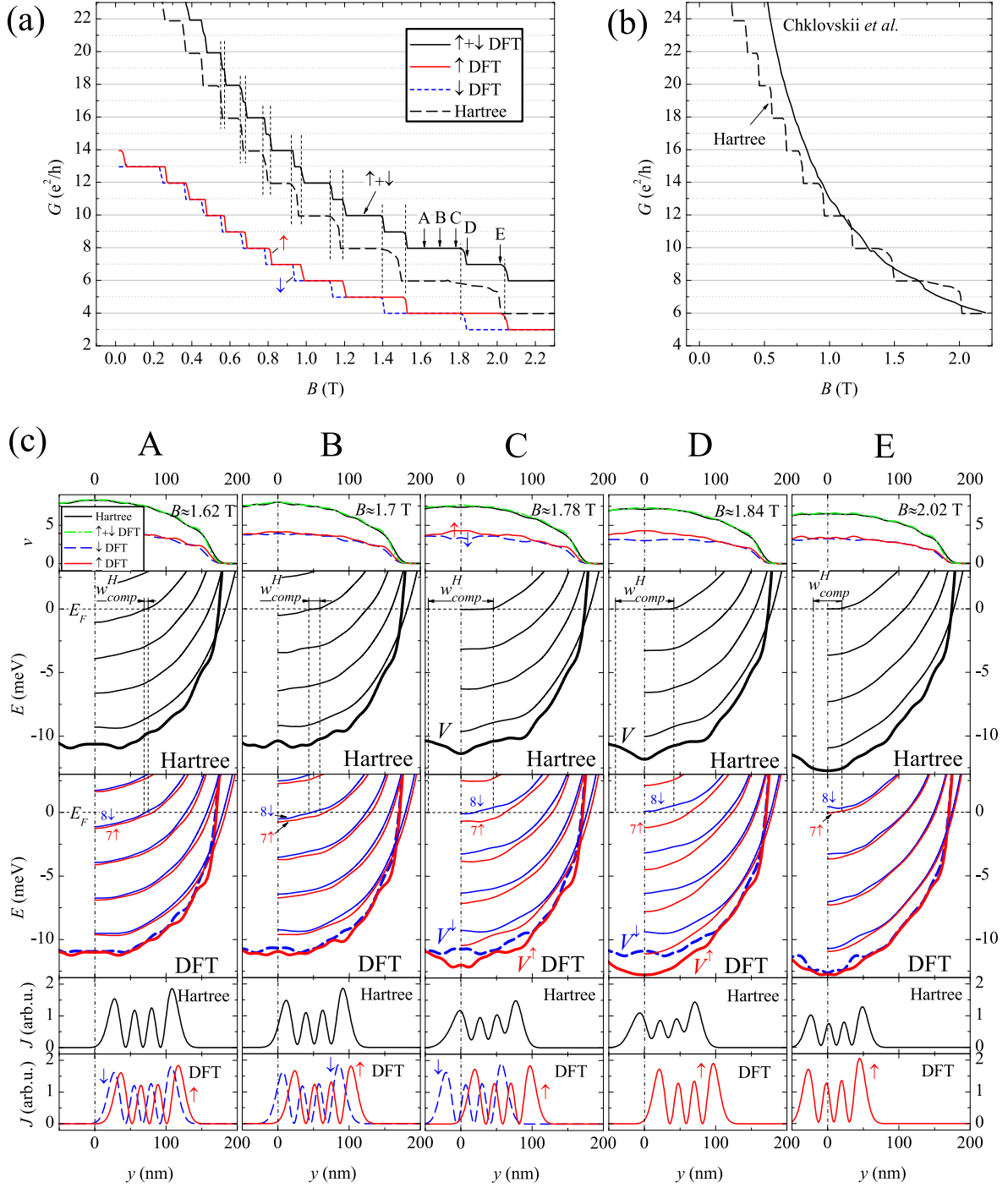} 
\caption{(a) Conductance of the quantum wire calculated within the spin DFT
and in the Hartree approximation (the later is shifted by $-2e^{2}/h$ for
clarity). The vertical lines are drawn to emphasize that the width of the
odd conductance steps in the spin DFT calculations is equal to the width of
the transition intervals between the conductance steps in the Hartree
calculations. (b) Comparison of between the Hartree magnetoconductance and
the magnetoconductance calculated according to the Chklovskii \textit{et al.}
conjecture $G_{Ch}=\frac{2e^2}{h}\protect\nu (0)$. (c) Evolution of the
magnetosubband structure in the interval $6e^{2}/h<G<8e^{2}/h$ (second and
third row). Fat solid lines indicate the total confining potential. The
first row shows the electron density profiles (local filling factors) $%
\protect\nu(y)=n(y)/n_{B}$ $(n_{B}=eB/h)$. Two lower rows show the current
densities for the last two subbands ($N=7,8$). Parameters of the wire are
the same as in Fig. \protect\ref{fig:structure}, $T=100$mK.}
\label{fig:G+subbands}
\end{figure*}
Figure \ref{fig:G+subbands}(a),(b) shows the magnetoconductance of a
representative wire with the effective width $w\approx 350$ nm and the
electron density in the center of the wire $n(0)=3.2\cdot 10^{15}$ m$^{-2}$
calculated within the Hartree and the spin DFT approximations. The Hartree
magnetoconductance shows the plateaus quantized in units of 2$e^{2}/h$
separated by transition regions whose width grows as the magnetic field is
increased. For large fields the width of the transition regions is
comparable or can even exceed the width of the neighboring plateaus. For low
fields, $B\lesssim B_{\text{crit}},$ the width of the transition regions
practically shrinks to zero; for the quantum wire at hand this critical
field is $B_{\text{crit}}\sim 0.6$T, corresponding to the subband index $%
N\approx 17,$ see Fig. \ref{fig:G+subbands}(a). Note that in a standard
one-electron picture of edge states the magnetoconductance of a clean wire
(without impurities) is strictly quantized in units of $2e^{2}/h$ (for
spinless electrons), with vanishing width of the transition regions.
Formation of the transition region between the plateaus is shown to be
related to development of the compressible strip in the middle of the wire%
\cite{ChklovskiiII}.

Let us now turn to the spin-resolved magnetoconductance calculated by the
spin DFT. \textit{The most striking feature of the wire magnetoconductance
is that the width of the odd conductance steps in the spin DFT calculations
is equal to the width of the transition intervals between the conductance
steps in the Hartree calculations}, see Fig. \ref{fig:G+subbands}(a). We
will demonstrate below that the characteristic features in the spin-resolved
conductance of the quantum wires calculated on the basis of the spin DFT
(including the dependence of the width of the odd plateaus on the magnetic
field and collapse of the odd plateaus at lower fields $B\lesssim B_{\text{%
crit}}$) can be understood from the analysis of the compressible strip
structure for spinless electrons and the corresponding magnetoconductance
and the magnetosubband structure evolution in the Hartree approximation.
\begin{figure}[tb]
\includegraphics[scale=1.0]{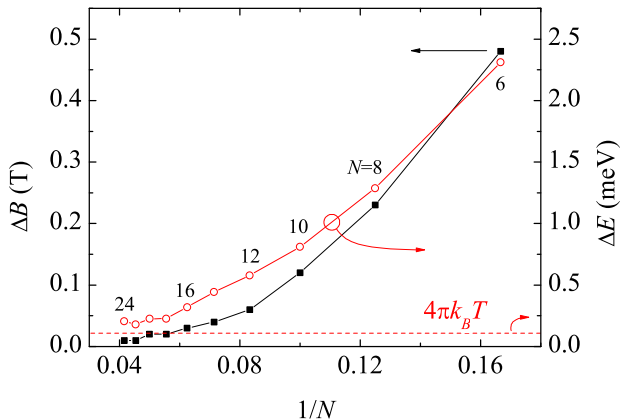} 
\caption{Solid line: The width of the transition regions between the Hartree
plateaus (which is equal to the width of the odd plateaus in the spin DFT
calculations) as a function of $1/N$, with $N$ being the subband number
(indicated in the plot). Thin line: The subband splitting in the wire center
$\Delta E=E$ (the definition of $\Delta E$ is illustrated in Fig. \protect
\ref{fig:magnetosubbands}). Parameters of the wire are the same as in Fig.
\protect\ref{fig:structure}, $T=100$mK. }
\label{fig:width}
\end{figure}

Before we proceed to the analysis of the evolution of the magnetosubband
structure, it is instrumental to outline how the exchange interaction
induces the spin splitting (see for details Refs. \onlinecite%
{Ihnatsenka,Ihnatsenka2}). The spin splitting is most pronounced in the
compressible strips. Indeed, the Hartree compressible strips are formed for
partially occupied states in the vicinity of $E_{F}$ when the Fermi-Dirac
occupation $f_{FD}<1$ (i.e. in the window $|E-E_{F}|<2\pi kT$). When the
states are partially occupied, the system behaves like a metal, where the
electrons can easily readjust their density to screen the external
potential. It is important to stress that the compressible strips, being
partially occupied, allow for different population of the spin-up and
spin-down states. In the DFT calculations, this population difference
(triggered by the Zeeman splitting) is strongly enhanced by the exchange
interaction. This leads to the lifting of the subband degeneracy and to the
spatial separation between the spin-up and spin-down states.

\subsection{The magnetosubband structure and the magnetoconductance}

In order to get insight into evolution of the odd conductance plateaus let
us inspect the Hartree and the spin DFT magnetosubband structure. Let us,
for example, concentrate at the field region where $6e^{2}/h<G<8e^{2}/h$,
i.e. when the magnetoconductance clearly shows the spin splitting. The
magnetosubband structure for several representative fields in this region is
shown in Fig. \ref{fig:G+subbands}(c). In all subsequent discussions we will
focus on the two highest subbands (in this case $N=8$ and $N=7$), because
the depopulation of these two subbands determines the features in the
conductance steps (note that all the remaining subbands are fully filled).
At $B=1.62$T the subbands $N=7,8$ are fully occupied and thus the total
conductance $G=8e^{2}/h$. The Hartree calculations for spinless electrons
show the presence of a narrow compressible strip of the width $w_{comp}^{H}$%
, see Fig. \ref{fig:G+subbands} A. When the exchange interaction is
included, the subbands split which leads to the spatial separation between
the spin-up and spin-down states. (The spatial spin separation due to the
suppression of the Hartree compressible strips was discussed in detail in
Ref. \onlinecite{Ihnatsenka2}).

When magnetic field is increased the subbands are pushed up in energy (see
Fig. \ref{fig:G+subbands} B; $B=1.7$T). The compressible strip in the
Hartree calculations becomes wider (because the confinement is smoother in
the wire center), and it moves closer to the center of the wire. The
exchange interaction quenches the compressible strip causing the splitting
of the spin-up and spin-down subbands. However, despite of the lifting of
the spin degeneracy the subband bottoms are still below $E_{F}$ at the wire
center. Because of this, the two spin split subbands $N=7,8$ remain fully
(and equally) populated and the conductance remains on the plateau $%
G=8e^{2}/h$.

When the magnetic field is increased to $B=1.78$T (Fig. \ref{fig:G+subbands}
C), the Hartree compressible strip reaches the middle of the wire. This
means that the subbands become partially occupied because their bottoms are
now within the window $|E-E_{F}|<2\pi kT$ (where $f_{FD}<1$). As a result,
the conductance of the spinless Hartree electrons starts to decrease and the
transition region between the plateaus starts to form. With further increase
of the magnetic field the Hartree compressible strip in the middle of the
wire shrinks, and at $B\approx 2.05$T the subbands depopulate completely, as
they are pushed above the window $|E-E_{F}|=2\pi kT$. We conclude the
discussion of the evolution of the Hartree subbands by re-emphasizing the
fact that the transition between the conductance steps starts when the
compressible strip reaches the center of the wire and it ends when the
compressible strip disappears and two highest (spin-degenerate)
magnetosubband are pushed above $E_{F}$. Note that even though this
discussion was focused on the transition between $G=8e^{2}/h$ and $6e^{2}/h$%
-plateaus in the Hartree conductance, the same scenario of the Hartree
subband depopulation holds for all other subbands.

Let us now examine how the exchange interaction affects the transition
region between the Hartree plateaus $G=8e^{2}/h$ and $6e^{2}/h$. Similarly
to the cases of lower fields discussed above (Fig. \pageref{fig:G+subbands}
A, B), the exchange interaction causes the subband repulsion and the spatial
spin separation of the wavefunctions (the latter being equal to the width of
the Hartree compressible strips, see the lower panel of Fig. \ref%
{fig:G+subbands}(c)). For $B=1.78$T the Hartree compressible strip covers
the central part of the wire. As a result, the bottom of the higher energy
(spin-down $N=8$) subband is situated within the window $|E-E_{F}|<2\pi kT$
(and thus this subband is only partially populated), whereas $N=7$
(spin-down) subband is pushed below $E_{F}$ and thus remains fully populated
(Fig. 4 C, spin DFT calculations). Thus, at $B=1.78$T the transition to the
odd plateau $G=7e^{2}/h$ starts to form. The exchange interaction keeps 8th
and 7th subbands separated such that with further increase of the magnetic
field 8th subband becomes quickly depopulated while 7th subband remains
fully occupied with its bottom being below $E_{F}$, see Fig. 4 D, E ($B=1.84$%
T and $B=2.05$T). The above field interval (i.e. $1.84\text{T}<B<2.05$T)
corresponds to the odd step in the magnetoconductance. Finally, at $B=2.05$T
(i.e. at the same field when the corresponding Hartree subbands depopulate),
the bottom of 7th subband is pushed above $E_{F}$ (to be more precise, above
$E_{F}+2\pi kT$) and 7th plateau in the conductance disappears.

To summarize the discussion presented in this section, the formation of the
odd magnetoconductance plateaus due to the exchange interaction can be
traced to the formation of the compressible strips in the center of the wire
in the case of the spinless electrons. The exchange interaction lifts the
spin degeneracy such that the bottom of the highest (even) subband remains
pinned to $E_{F}$, whereas the bottom of the highest odd subband remains
below $E_{E}$. As a result, the odd plateaus (whose width is equal to the
width of the transition regions between the Hartree plateaus) develop in the
magnetoconductance.

Note that the analytical solution to the electrostatic problem of the electron density
distribution in a quantum wire for spinless electrons has been obtained by Chklovskii,
Matveev and Shklovski \cite{ChklovskiiII} for the high magnetic field regime (when only a
few lower Landau levels are occupied, $N\lesssim 2\sim 4$). [A good agreement with with
the analytical results of Chklovskii \textit{et al. } has been reported by Oh and
Gerhardts within the self-consistent Thomas-Fermi calculations \cite{Oh}]. Chklovskii
\textit{et al.}\cite{ChklovskiiII} have also discussed the magnetoconductance of the
quantum wire. They found that in a realistic quantum wire the conductance plateaus are
practically absent, i.e. the conductance is not quantized (see Fig. 5 in Ref.
\onlinecite{ChklovskiiII}). This conclusion is in obvious disagreement with the
experimental results showing pronounced plateaus in the two-terminal magnetoconductance
at integer values of $e^{2}/h.$ \cite{BvH,Wrobel,experiment}. They attributed this
discrepancy to the presence of disorder in the channel. We however, have demonstrated
above that even in an ideal clean channel (without disorder) the conductance shows the
pronounced quantization with wide plateaus and sharp rises. The reason of the discrepancy
we instead attribute to the conjecture used by Chklovski\textit{et al.} that the
ballistic conductance is given by the filling factor in the middle of the wire,
$G_{Ch}=\frac{2e^{2}}{h}\nu (0)$. Our quantum mechanical calculations show that this
conjecture is not justified. Indeed, the comparison of the conductance calculated
according to the Chklovskii \textit{et al. }conjecture with the magnetoconductance
calculated according to Eq. (\ref{G}) (see Fig. \ref{fig:G+subbands}(b)) shows that
$G_{Ch}$ reproduces the overall decrease of the conductance rather well, but does not at
all recover the steps in the conductance related to the subband depopulation (see also
Ref. \onlinecite{Lier} for a related discussion). Our results thus indicate that while
electrostatic and Thomas-Fermi-type approaches can be very successful in the description
of the electron density and the structure of the compressible and incompressible strips
for the spinless electrons, an accurate description of the magnetoconductance requires
detailed quantum mechanical information for the wave functions and the currents
densities.

\subsection{Collapse of the odd magnetoconductance plateaus at lower fields}

When magnetic field is increased, the width of the transition regions
between the Hartree plateaus (which is equal to the width of the odd
plateaus in the spin resolved magnetoconductance), $\Delta B$, is also
gradually increases, see Figs \ref{fig:G+subbands},\ref{fig:width}. We
attribute this increase of $\Delta B$ to the effect of the enhanced electron
screening due to the evolution of the compressible strip in the middle of
the wire. Indeed, the transition regions between the Hartree plateaus are
related to the depopulation of the highest (spin-degenerate) subbands
forming the compressible strip in the center. In high magnetic field each
subband (representing a Landau level) accommodates the same number of
electrons, such that the density of the electrons in the highest subband is
proportional to $\sim 1/N$. Thus, one can expect that the width of the
compressible strip in the middle of the wire $w_{comp}$ and hence the width $%
\Delta B$ grow as $B$ increases (note that $B\sim 1/N)$. Figures \ref%
{fig:width} - \ref{fig:currents} illustrating the magnetic field dependence
of $\Delta B$ and $w_{comp}$ confirms this expectation. Note that $\Delta B$
shows a nonlinear dependence on $1/N$. That is, for low fields, $B\lesssim
B_{\text{crit}},$ the width $\Delta B$ rapidly decreases when $B$ decreases,
such that the odd plateaus are no longer seen in the magnetoconductance.

\begin{figure}[tb]
\includegraphics[scale=0.7]{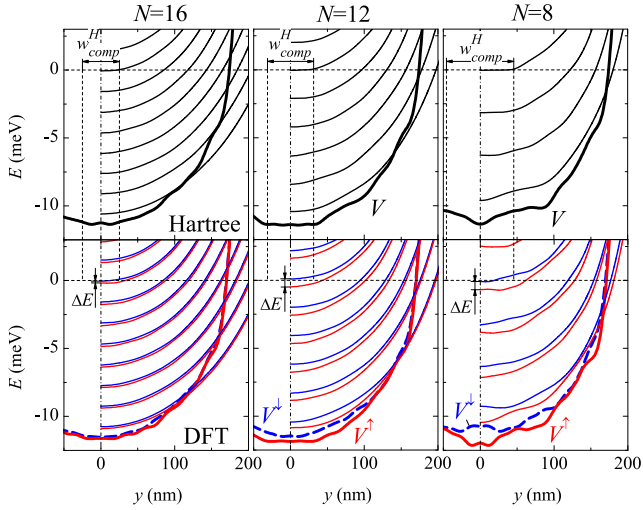} 
\caption{The magnetosubband structure of a quantum wire within the Hartree
and the spin-DFT approximation for different number of occupied subbands $N$%
. Fat solid lines indicate the total confining potential. [For all cases the
magnetic field is chosen such that the Hartree compressible strips in the
middle of the wire have a maximal width]. $\Delta E$ shows the subband
splitting in the center of the wire. Parameters of the wire are the same as
in Fig. \protect\ref{fig:structure}, $T=100$mK. }
\label{fig:magnetosubbands}
\end{figure}

Let us now concentrate on this feature of magnetoconductance in more detail.
Figure \ref{fig:G+subbands} shows the spin-resolved magnetoconductance $%
G^{\uparrow }$ and $G^{\downarrow }$ . It is worth to stress that the spin
degeneracy remains lifted even for fields smaller than $B_{crit}$. The total
conductance $G=G^{\uparrow }+G^{\downarrow }$ does not however exhibits the
odd plateaus for $B<B_{crit}$ because the strength of the exchange splitting
becomes comparable to the thermal broadening of the plateaus. This is
illustrated in Fig. \ref{fig:width} that shows the dependence of the subband
splitting $\Delta E$ in the center of the wire on the magnetic field and its
comparison to the energy window $4\pi kT$ (where the derivative of the
Fermi-Dirac distribution function is distinct from zero). [The definition of
the subband splitting $\Delta E$ is outlined in Fig. \ref%
{fig:magnetosubbands}]. It is also worth pointing out that in accordance to
the previous discussion, $\Delta E$ and $\Delta B$ exhibit similar behavior
as a function of magnetic field, see Fig. \ref{fig:width}.

\begin{figure}[tb]
\includegraphics[scale=1.0]{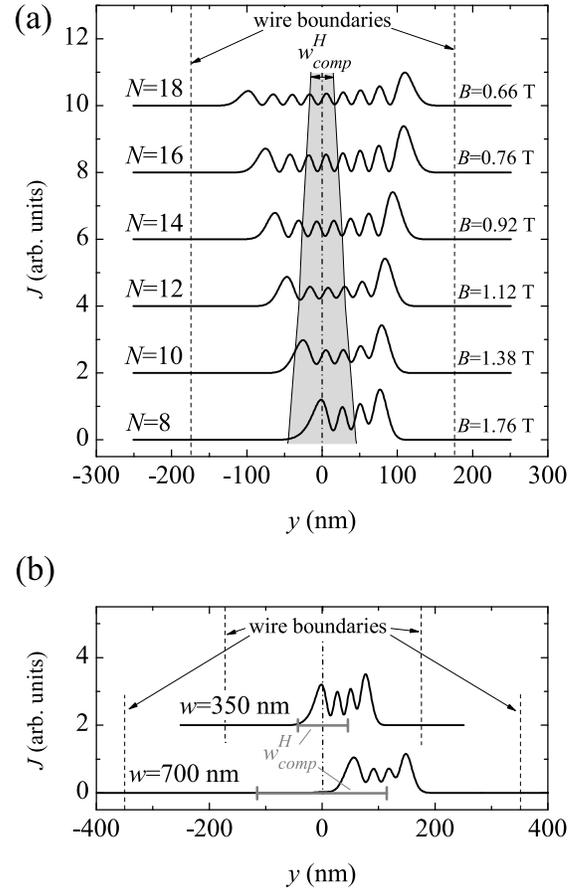} 
\caption{(a) The current densities for the highest occupied Hartree subband
compared to the maximal width of the compressible strip in the middle of the
wire for different magnetic fields. $N$ indicates the subband number (see
Fig. \protect\ref{fig:magnetosubbands} for the corresponding magnetosubband
structure). (b)The current densities for the highest occupied Hartree
subband ($N=8$) compared to the maximal width of the compressible strip in
the wires with the effective widths of $w=350$ nm and $w=700$ nm. Parameters
of the wire are the same as in Fig. \protect\ref{fig:structure}, parameters
of the second wire are indicated in Fig. \protect\ref{fig:comparison}; $%
T=100 $mK. }
\label{fig:currents}
\end{figure}

In order to understand the nonlinear behavior of $\Delta B$ leading to
quenching of the odd magnetoconductance plateaus at low field we examine the
wave functions and the current density distributions. Figure \ref%
{fig:currents} (a) shows the current density for the Hartree subbands $%
N=8-18 $ along with the maximal width of the Hartree compressible strip $%
w_{comp}$ in the middle of the wire. The extend of the wave function for the
highest $N$th subband, $\langle \psi _{N}\rangle \sim \sqrt{N}l_{B},$ ($%
l_{B} $ is the magnetic length) gradually increases when the magnetic field
is lowered\cite{Davies_book}. Note that $\langle \psi _{N}\rangle $ is
larger than the width of the compressible strip $w_{comp}$ already for $N=8$%
. When the extend of the wave function exceeds the width of the compressible
strip, the ability of the system to screen the external potential is greatly
reduced because the wave function can be shifted within the distance not
exceeding the width of the compressible strip $w_{comp}$. Thus, the smaller
the ratio $w_{comp}/\langle \psi _{N}\rangle$ is, the weaker is the effect
of the redistribution of the electron density required to screen the
external potential. This reduced screening efficiency for lower fields (when
$w_{comp}/\langle \psi _{N}\rangle \ll 1)$ translates into the suppressed
exchange splitting and thus to disappearance of the odd magnetoconductance
plateaus.

Note that the extend of the wave function $\langle \psi _{N}\rangle $ for a
given subband number $N$ (or for a given magnetic field) is not particularly
sensitive to the wire width $w$ (at least in the regime when the cyclotron
radius $r_{c}<w$). At the same time, the maximum width of the compressible
strip increases with increase of the wire width. This is illustrated in Fig. %
\ref{fig:currents}(b) that shows the current density distribution and $%
w_{comp}$ for two quantum wires of the with $w=350$ nm and $700$ nm for the
case of $N=6$ occupied subbands. [Note that in the bulk limit (i.e. for the
edge of the 2DEG) the compressible strip covers the semi-infinite space,
such that regardless of the subband number, $w_{comp}/\langle \psi
_{N}\rangle \gg 1$]. Therefore, for the given $N$ (or magnetic field $B$)
the ratio $w_{comp}/\langle \psi _{N}\rangle $ is larger in a wider wire and
therefore the screening efficiency is higher. One can therefore expect that
in a wider wire the magnetosubband spin splitting due to exchange
interaction leading to the appearance of the odd magnetoconductance plateaus
would manifest itself for larger subband numbers (lower fields). Our
calculations show that this is indeed the case. For example, in a wire with $%
w=350$ nm the odd plateaus become discernible for the subband index $N=17$ ($%
B_{\text{crit}}\sim 0.6$T), whereas for the wire with $w=700$ nm the last
odd plateau is seen for $N\sim 19-21$ ($B_{\text{crit}}\sim 0.4$T), c.f.
Figs. \ref{fig:G+subbands} and \ref{fig:comparison}

\section{Comparison to the experiment}

We have performed magnetotransport calculations for several quantum wires
with effective widths in the range 200-700 nm and the electron densities $%
1.5-3.5\cdot 10^{15}$m$^{-2}.$ All the wires exhibit the same behavior
described in detail in previous section.

\begin{figure}[tb]
\includegraphics[scale=1.0]{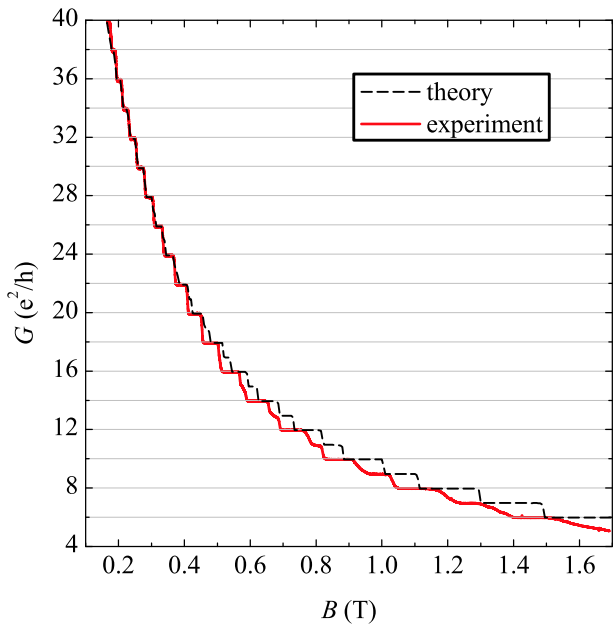} 
\caption{Comparison between the calculated magnetoconductance of the wire
with the effective width of $w=700$nm and the experimental magnetoconductance%
\protect\cite{experiment}. $T=100$mK. }
\label{fig:comparison}
\end{figure}

A detailed comparison of our calculations with the experimental
magnetoconductance\cite{experiment} for a some representative quantum wire
is shown in Fig. \ref{fig:comparison}. The width of the wire is estimated to
be $\sim 680$ nm, and the sheet electron density in the bulk $%
n_{s}=2.15\cdot 10^{15}$m$^{-2}$.\cite{experiment}. The theoretical
magnetoconductance shown in Fig. \ref{fig:comparison} is calculated for the
electrostatic confining potential with the parameters $V_{0}=-0.4$ eV, $%
\hbar \omega _{0}=1.91$ meV (giving the effective wire width $w=700$ nm and
the electron density in the wire center $n=2.2\cdot 10^{15}$m$^{-2}$). In
the both experimental and simulated wires the electrons is situated at the
distance $\sim 200$ nm below the surface. The comparison to the subband
depopulation in the experimental structure demonstrates that such a choice
of the parameters provides a satisfactory approximation for the actual
confining potential. We stress here that a magnetic field dependence of the
subband depopulation can be described by the one-particle Schr\"{o}dinger
equation (for a given confining potential)\cite{Berggren}, whereas our main
focus here is the electron interaction effects leading to formation of the
odd steps in the magnetoconductance due to the exchange interaction. The
comparison of the calculated and the experimental curves demonstrates a good
quantitative agreement between the widths of the odd (as well as even)
plateaus in the calculated and the experimental magnetoconductance. The
calculations also provide a reasonably close estimation of the subband index
corresponding to the last resolved odd plateau in the magnetoconductance, $%
N\sim 19-21$, whereas the corresponding experimental value is $N=15.$

It would be unreasonable to expect an exact agreement between the theory and
the experiment. There are several factors that have not been taken into
account in the theoretical modelling. We list below some of them.

(a) The experiment\cite{experiment} is performed in the QPC geometry whereas
our calculations are done for an infinite quantum wire. In the edge state
transport regime considered here with $N\gg 1$ this is not expected to be a
source of significant discrepancy between the theoretical predictions and
the experiment. Nevertheless, the effect of the QPC geometry on the
magnetoconductance remains to be seen.

(b) The calculations are performed for ideal clean wires without impurities.
In a one electron description the transition regions between the
magnetoconductance plateaus are the step functions with a zero width. A
random impurity potential is known to lead to the smooth transition regions
of a finite width even in the one-electron picture\cite{Ando_1994}. A
comparison of the magnetoconductance traces in Fig. \ref{fig:comparison}
clearly shows that the the transition regions seen in the experiment are
significantly wider than the theoretical ones. [We remind that a finite
width of the transition regions in the theoretical magnetoconductance is due
to the formation of the compressible strips in the middle of the wire]. We
attribute the difference in the widths of the calculated and experimental
transition regions to the effect of the disorder that has not been included
in the model. Note the experimental magnetoconductance traces for narrower
QPCs show the conductance fluctuations in the transition regions\cite%
{experiment}, which is a clear manifestation of the disorder potential due
to impurities\cite{disorder}. Note that the presence of disorder can also
lead to the destruction of the exchange enhancement of the $g$-factor and
thus to the collapse of the spin splitting (i.e. to the suppression of the
odd plateaus for $N>N_{c}$)\cite{Fogler}. This effect does not seem to be
relevant to the experiment because the spin splitting, being suppressed in
the narrow structures (QPCs) for larger $N$, is still clearly seen in the
bulk Hall measurements.

(c) In our calculations we assumed that an electron motion is confined to a
two-dimensional plane, which is a good approximation for heterostructures
where the electrons are localized on the interface between GaAs/AlGaAs. In
the experimental structures\cite{experiment} the electrons are confined in a
quantum well that is populated by donors situated on both sides from the
well. An accurate description of this geometry might require accounting in
the Schr\"{o}dinger equation the electron motion in the direction
perpendicular to the interface.

Finally, we notice that while we compared our calculations with the
experimental conductance of one representative wire, our spin-DFT
calculations qualitatively reproduce all the features observed in other
samples. This includes the dependence of of the width of the odd plateaus $%
\Delta B$ on the magnetic field shown in Fig. \ref{fig:width}. It is also worth to stress
that the theory confirms (and explains) the experimental finding that in wider wires the
collapse of the odd plateaus occurs at lower fields. We however are not in position to
fit all the experimental data. This is because that this task would require a detailed
knowledge of the bare confining electrostatic potential $V_{conf}$ due to the gates and
the donor layers, Eq. (\ref{V_conf}). Indeed, the bare electrostatic confinement
determines the total self-consistent confining potential $V^{\sigma }(y)$, which, in
turn, determines the depopulation of the magnetosubbands (i.e. the dependence of the
subband number $N$ on $B$)\cite{BvH,Berggren}. We are not in a position to perform a
systematic search for the parameters of $V_{conf}$ giving rise to the $B$-dependence of
the subband depopulation consistent with each experimental magnetoconductance trace. This
is simply because of a computation burden related to this task: each point on the
magnetoconductance plot requires up to one hour of a processor time. [Note that a
calculation of the electrostatic confinement and the self-consistent potential starting
from the layout of the actual heterostructure of Ref. \onlinecite{experiment} represents
a separate task which is outside a scope of the present study].

\section{Conclusion}

In this paper we provide a systematic quantitative description of the
magnetoconductance of the split-gate quantum wires focusing on the formation
and evolution of the odd conductance plateaus. In order to calculate the
electron density, magnetosubband structure and the magnetoconductance we
utilize the self-consistent Greens function technique combined with the spin
density functional theory\cite{Ihnatsenka}.

We start our analysis with the case of spinless electrons in the Hartree
approximation (disregarding the exchange and correlation interactions). The
calculated Hartree magnetoconductance shows the plateaus quantized in units
of 2$e^{2}/h$ separated by transition regions whose width grows as the
magnetic field is increased. The transition regions are attributed to the
formation of the compressible strips in the middle of the wire occupied by
electrons belonging to the highest (spin-degenerate) subband. In agreement
with experiments, the width of the transition regions for large fields is
comparable to the width of the neighboring plateaus. This is in contrast to
both the one-electron description where the conductance shows the step-like
behavior with the rises between the plateaus of zero width, as well as to
the electrostatic theory of Chklovskii \textit{et al.}\cite{ChklovskiiII}
where the magnetoconductance exhibits narrow plateaus of negligible width
separated by much broader transition regions where the conductance is not
quantized.

Accounting for the exchange and correlation interactions within the spin DFT
leads to the lifting of the spin degeneracy and formation of the
spin-resolved plateaus at odd values of $e^{2}/h.$ The most striking feature
of the magnetoconductance is that the width of the odd conductance steps in
the spin DFT calculation is equal to the width of the transition intervals
between the conductance steps in the Hartree calculations. This is because
the transition intervals in the Hartree magnetoconductance correspond to the
formation of the compressible strip in the middle of the wire. At the same
time, in the compressible strip in the center of the wire the states are
only partially occupied. As a result, the exchange interaction enhances the
difference in the spin-up and spin-down population, which leads to the
lifting of the subband spin degeneracy and formation of the odd conductance
plateaus.

In agrement with the experimental results\cite{experiment}, we find that the
width of the odd magnetoconductance plateaus gradually decreases with
decrease of the magnetic field. For lower fields $B<B_{\text{crit}}$, the
odd plateaus rapidly disappear such that the magnetoconductance shows the
quantization in units of $2e^{2}/h.$ The wider the wire, the lower the
critical field $B_{\text{crit}}$ corresponding to the disappearance of the
last resolved odd plateau. We attribute this effect to the reduced screening
efficiency in the confined (wire) geometry when the width of the
compressible strip in the center becomes much smaller than the extend of the
wave function. This, in turn, leads to the suppressed exchange splitting and
collapse of the odd magnetoconductance steps.

A detailed comparison to the experimental data\cite{experiment} (see Fig. %
\ref{fig:comparison}) demonstrates that the spin-DFT calculations reproduce
not only qualitatively, but rather quantitatively all the features observed
in experiment. This includes the dependence of the width of the odd and even
plateaus on the magnetic field as well as the estimation of the subband
index corresponding to the last resolved odd plateau in the
magnetoconductance. The experiment however shows wider rises between the
transitions plateaus in comparison to the calculated ones. We attribute this
difference to the effect of smooth potential due to remote donors that has
not been accounted for in our calculations (performed for clean
disorder-free wires). Despite of this discrepancy, the overall good
agreement between the theory and experiment makes it possible to conclude
that the spin DFT approach represents the powerful tool to study large
realistic quantum Hall systems containing hundreds or thousands of
electrons, providing detailed and reliable microscopic information on
wavefunctions, electron densities and currents as well as the conductance.


\begin{acknowledgments}
We are thankful to C. M. Marcus for drawing our attention to the current problem. We also
appreciate discussions and correspondence with the authors of the Ref.
\onlinecite{experiment} (I. Radu, C. M. Marcus, M. Kastner) and we are thankful for their
kind permission to use their experimental data prior publication. We acknowledge access
to computational facilities of the National Supercomputer Center (Link\"{o}ping) provided
through SNIC.
\end{acknowledgments}


\end{document}